\documentclass[11pt]{article}

\usepackage[T1]{fontenc}
\usepackage[utf8]{inputenc} 
\usepackage[english]{babel}
\addto\captionsenglish{%
}
\addto\captionsitalian{%
}

\usepackage[a4paper,margin=28mm]{geometry}
\usepackage{lmodern}
\usepackage{microtype}
\usepackage{setspace}
\usepackage{booktabs}
\usepackage{tabularx}
\usepackage{ragged2e}

\usepackage{amsmath,amssymb}
\usepackage{algorithm}
\usepackage{algpseudocode}

\usepackage{tikz}
\usetikzlibrary{arrows.meta, positioning}

\usepackage{titlesec}

\titleformat{\section}
{\normalfont\huge\bfseries}
{\thesection}{1em}{}
\titlespacing*{\section}
{0pt}{3ex plus 1.2ex minus .3ex}{1.8ex plus .3ex}

\titleformat{\subsection}
{\normalfont\large\bfseries}
{\thesubsection}{1em}{}
\titlespacing*{\subsection}
{0pt}{2ex plus 1ex minus .2ex}{1ex plus .2ex}

\titleformat{\subsubsection}
{\normalfont\normalsize\bfseries}
{\thesubsubsection}{0.8em}{}
\titlespacing*{\subsubsection}
{0pt}{1.5ex plus .8ex minus .2ex}{0.8ex plus .2ex}

\usepackage[hidelinks]{hyperref}

\title{A Real-Options-Aware Multi-Criteria Framework for Ex-Ante Real Estate Redevelopment Use Selection}

\author{
	Roberto Garrone\\[0.5ex]
	\small
	University of Salford, Salford, UK\\
	\small
	University Sapienza Unitelma, Rome, Italy\\
	\small
	University of Milan-Bicocca, Milan, Italy\\
	\small
	\texttt{roberto.garrone@unimib.it}
}

\date{\today}

\begin{document}
	\maketitle
	
\begin{abstract}
	A growing share of the existing real estate stock exhibits persistent underperformance that can no longer be explained by cyclical market phases or inadequate maintenance alone. In many cases, technically recoverable assets located in non-marginal contexts fail to generate economic value consistent with the capital immobilized. This condition reflects a structural misalignment between intended use and effective demand rather than episodic market weakness, and calls for a decision framework capable of integrating value, risk, complexity, and irreversibility in strategic use selection.
	
This study proposes a decision-analytic framework for the ex-ante selection of intended
use in real estate redevelopment processes. The framework integrates real-options logic
on irreversibility and managerial flexibility with a multi-criteria decision-analysis
structure, enabling comparative evaluation of expected economic value, market and
operational risk, technical and managerial complexity, and time-to-income.

	By treating redevelopment primarily as a problem of strategic option selection rather than design or financial optimization, the framework operationalizes option value preservation through disciplined ex-ante screening. Illustrative cases demonstrate how this integration of real options reasoning and MCDA reduces over-complexification and misalignment across different asset types and urban contexts.
\end{abstract}


\section{Introduction}

A growing share of the existing real estate stock exhibits persistent underperformance that can no longer be adequately explained by cyclical market phases or by deficiencies in routine maintenance alone. In many cases, technically recoverable assets located in non-marginal urban contexts fail to generate economic value commensurate with the capital they immobilize. This pattern is increasingly structural rather than episodic, and cannot be satisfactorily attributed to generic categories such as sectoral downturns or temporary adverse economic conditions.

In a large subset of these cases, underperformance reflects a misalignment between the asset’s intended use and the effective demand that its surrounding context is able to sustain. Buildings designed for work, living, or hospitality models that have become functionally outdated often continue to be operated according to economic logics that no longer correspond to evolving demographic, social, and organisational conditions. Over time, this mismatch tends to manifest in persistent vacancy, rent compression, erosion of asset value, and heightened operational and financial risk.

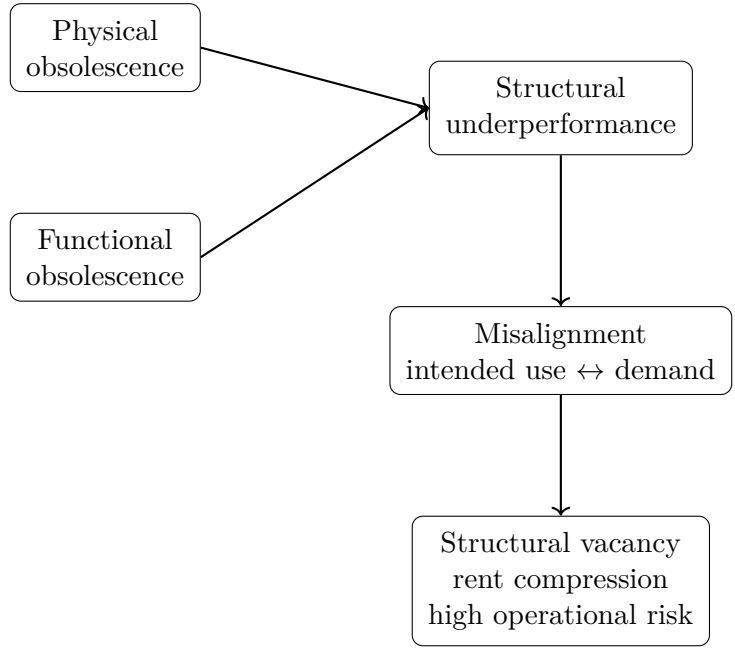
\begin{figure}[t]
	\centering
	\begin{tikzpicture}[
		node distance=1.6cm,
		every node/.style={align=center},
		box/.style={draw, rounded corners, inner sep=6pt},
		arrow/.style={->, thick}
		]
		
		\node[box] (physical) {Physical\\obsolescence};
		\node[box, below=of physical] (functional) {Functional\\obsolescence};
		
		\node[box, right=3cm of physical, yshift=-0.8cm] (underperf)
		{Structural\\underperformance};
		
		\node[box, below=2cm of underperf] (misalign)
		{Misalignment\\intended use $\leftrightarrow$ demand};
		
		\node[box, below=of misalign] (effects)
		{Structural vacancy\\
			rent compression\\
			high operational risk};
		
		\draw[arrow] (physical.east) -- (underperf.west);
		\draw[arrow] (functional.east) -- (underperf.west);
		\draw[arrow] (underperf) -- (misalign);
		\draw[arrow] (misalign) -- (effects);
		
	\end{tikzpicture}
	\caption{Conceptual scheme of structural underperformance in existing real estate assets}
	\label{fig:obsolescence-underperformance}
\end{figure}

Figure~\ref{fig:obsolescence-underperformance} schematically represents this process, linking physical and functional obsolescence to structural underperformance through the mediating role of use--demand misalignment.

A central implication of this diagnosis is that redevelopment is more appropriately understood as a decision problem than as a purely construction- or finance-driven optimisation task. In particular, the initial choice of intended use exerts a disproportionate influence on both risk exposure and value creation over the lifecycle of a redevelopment project. Once decisions concerning use, scale, and operating model have been fixed, irreversibility increases markedly, reducing the scope for subsequent correction and diminishing the option value embedded in the asset.

Despite this, early redevelopment choices are often guided by decontextualised market benchmarks, EUR/m$^{2}$ comparisons across non-homogeneous assets, or trend-driven narratives that are insufficiently grounded in verifiable, paying demand. Business plans derived from such premises may appear internally coherent, yet frequently prove fragile in execution because they inadequately account for operational risk, managerial intensity, regulatory uncertainty, and time-to-income dynamics.

Against this background, the objective of this study is to develop an operational and replicable framework for the \textit{ex ante} selection of intended use in the redevelopment of assets with unrealised potential. The framework is deliberately positioned upstream of feasibility studies and detailed financial modelling. Its purpose is not to identify universally ``optimal'' real estate formats, nor to promote specific typologies as generic solutions, but to reduce the probability of strategic misalignment before irreversible commitments of capital, time, and design resources are undertaken.

To this end, the proposed approach adopts a comparative, multi-criteria perspective structured around four core dimensions: expected economic value, proxied through conservative and relative Net Present Value estimates; risk, decomposed into market and operational components; complexity, articulated across technical, regulatory, and managerial dimensions; and time-to-income, reflecting both financial sustainability and consistency with the decision-maker’s investment horizon. These dimensions are integrated into a decision matrix designed as an \textit{ex ante} selection and exclusion device rather than as a point-forecasting instrument.

The primary contribution of this study lies not in proposing new real estate formats, but in providing a disciplined method for strategic use selection under structural uncertainty. In a context characterised by increasingly selective capital, extended authorisation processes, and fragmented demand, reducing \textit{ex ante} decision error becomes a key mechanism for protecting value. By making trade-offs between value, risk, and complexity explicit at the earliest stages of the redevelopment process, the framework aims to improve decision quality and preserve flexibility. It is intended for use by asset owners, developers, and investment committees at the screening stage, prior to feasibility analysis, and is not designed to generate executable designs or precise return forecasts.

This paper contributes to the strategic property management and redevelopment literature by providing a formal decision-support framework for ex-ante use selection, situated at the intersection of real options theory and multi-criteria decision analysis. While real options models have traditionally focused on pricing flexibility and timing under uncertainty \cite{dixitpindyck1994, trigeorgis1996, grenadier1995}, and MCDA has been widely applied to property and construction decisions \cite{zavadskas2011, banaitis2011, peldschus2008}, their integration at the stage of initial use selection in redevelopment remains limited. This study addresses that gap by operationalising irreversibility, complexity, and execution risk within a comparative MCDA framework.

\section{Literature Review}

\subsection{Real Options and Irreversible Investment Decisions}

Traditional investment appraisal techniques in real estate—such as return on investment
(ROI), yield, and internal rate of return (IRR)—are ill-suited for decisions characterised
by uncertainty, irreversibility, and long time-to-market horizons. These indicators are
typically static, ex-post measures that fail to account for the strategic value of
flexibility and the costs associated with premature commitment \cite{dixitpindyck1994,trigeorgis1996}. As a result, they provide
limited guidance in contexts where uncertainty is structural and decisions introduce
significant irreversibility.

Real Options Theory (ROT) offers a conceptual framework for addressing these limitations
by interpreting investment decisions as options that can be deferred, staged, abandoned,
or reconfigured as uncertainty resolves over time. Foundational contributions in this
literature formalised the economic value of waiting and managerial flexibility in
irreversible investments, demonstrating that the timing of commitment constitutes a
strategic variable in its own right \cite{mcdonaldsiegel1986,dixitpindyck1994,trigeorgis1996}. Within this framework, premature investment may destroy option value by eliminating
future flexibility.

Subsequent applications of real options reasoning to real estate have shown that land and
redevelopment opportunities embed significant option premiums, particularly in contexts
characterised by regulatory uncertainty, demand volatility, and long development cycles
\cite{titman1985,quigg1993,grenadier1995}. In this literature, development timing, phased
construction, and temporary underutilisation are often interpreted as rational responses
to irreversibility rather than as market failure.

A key insight emerging from this literature is that the primary source of risk in
redevelopment is not volatility per se, but irreversibility. Once capital is committed,
design choices are fixed, and regulatory approvals are obtained, the ability to adapt the
asset to changing conditions is sharply reduced. At this stage, errors in the initial
choice of use or operating model become difficult and costly to correct
\cite{dixitpindyck1994,grenadier1995}. In redevelopment and adaptive reuse contexts, irreversibility is not merely financial but also
functional and regulatory. Once a use is selected and spatially embedded through layout,
building services, and planning approvals, the option to reposition is sharply reduced even
when market conditions change. The adaptive reuse literature has repeatedly shown that
early functional choices create long-term path dependence in building performance
\cite{bullenlove2011,langston2008,remoywilkinson2012}.

However, classical real options models are predominantly financial in nature and are typically
grounded in stochastic price processes. As such, they are limited as early-stage screening
tools when non-financial dimensions such as managerial capability, regulatory complexity,
organisational burden, and operational risk dominate outcomes \cite{trigeorgis1996,amramkulatilaka1999}.

\subsection{Multi-Criteria Decision Analysis in Complex Investment Contexts}

Multi-Criteria Decision Analysis (MCDA) provides a complementary perspective by structuring
decisions involving multiple, often conflicting objectives. MCDA frameworks enable the
comparison of heterogeneous criteria—economic, technical, organisational, and temporal—
within a coherent evaluation structure, typically through processes of normalisation,
weighting, and aggregation \cite{keeneyraiffa1976,saaty1980,beltonstewart2002,figueira2005}.

In real estate and construction research, MCDA has been extensively used to support
development selection, property portfolio management, and urban regeneration under
multi-dimensional uncertainty. Applications include comparative evaluation of investment
alternatives, sustainability trade-offs, and redevelopment strategies
\cite{zavadskas2011,banaitis2011,peldschus2008}. These studies demonstrate the suitability of
MCDA for problems where economic performance, technical constraints, and managerial
requirements must be assessed jointly.

In investment and planning contexts, MCDA is valued less for point precision than for its
capacity to make trade-offs explicit, to support transparent decision processes, and to
integrate qualitative and quantitative dimensions under uncertainty
\cite{beltonstewart2002,figueira2005}.

Within real estate and construction research, MCDA has been widely used to evaluate
development alternatives, location choices, sustainability performance, and regeneration
strategies \cite{peldschus2008,zavadskas2011,banaitis2011}. These applications demonstrate
the suitability of MCDA for problems where value cannot be reduced to a single monetary
metric.

However, most real estate MCDA applications remain ex-post or design-oriented, implicitly
assuming that the set of development alternatives is already defined. As a result, MCDA is
typically applied after the strategic choice of intended use has been made, treating the
alternative set as exogenous rather than as a decision variable \cite{zavadskas2011,banaitis2011}.

Relatively few contributions focus on the ex-ante phase in which the choice of use is still
open, even though this stage exerts the greatest influence on long-term risk exposure and
value creation. The absence of structured ex-ante decision support contributes to the
persistence of strategic errors that cannot be remedied through later optimisation.

\subsection{Integrating Real Options and MCDA for Ex-Ante Use Selection}

Several authors have proposed integrating real options reasoning with MCDA in order to
address investment problems characterised by irreversibility, multiple objectives, and
managerial flexibility \cite{herathpark2002,amramkulatilaka1999}. In such hybrid frameworks,
real options theory informs the treatment of timing and commitment, while MCDA provides
the comparative structure needed to evaluate alternatives across heterogeneous dimensions.

This integration is particularly relevant in real estate redevelopment, where uncertainty
is often structural rather than stochastic, and where key drivers of performance—such as
regulatory delay, organisational capability, and operational intensity—cannot be
represented by price volatility alone \cite{dixitpindyck1994,grenadier1995,zavadskas2011}.

The framework proposed in this study builds on this integrated perspective. Unlike
option-pricing approaches, it treats irreversibility, complexity, and execution risk as
decision constraints rather than as stochastic payoffs. Rather than seeking to monetise
flexibility through option pricing, it operationalises real-options logic by preserving
reversibility through ex-ante exclusion of structurally incoherent alternatives,
penalisation of irreversible complexity, and explicit treatment of time-to-income. MCDA
provides the comparative engine through which asset–use configurations are evaluated
before capital, design, and regulatory commitments are made.


\section{Structural Context of the Real Estate Market}

\subsection{Functional Obsolescence of the Existing Stock}

The loss of performance affecting a significant portion of the existing real estate stock
cannot be explained solely through traditional categories such as physical deterioration
or inadequate maintenance. Increasingly, the primary cause lies in a less visible but more
structurally relevant phenomenon: functional obsolescence. This condition occurs when an
asset, while technically recoverable, is no longer suited to the uses, economic models, and
modes of utilization that characterize contemporary demand.

It is therefore essential to draw a clear distinction between physical obsolescence and
functional obsolescence. The former concerns the construction quality and technical systems
of the building and can generally be addressed through maintenance or refurbishment
interventions. The latter, by contrast, concerns the asset’s ability to perform an economic
function that is coherent with the market context in which it operates. An asset may be
physically efficient yet economically ineffective if its functional configuration is no
longer aligned with demand.

In this sense, many underperforming properties are not intrinsically “wrong” assets, but
rather poorly positioned ones. Their underperformance stems from factors such as an
inadequate scale, a location that has lost functional centrality, rigid layouts that are
difficult to adapt, or use constraints that prevent efficient reconfiguration. These
characteristics limit the asset’s ability to intercept solvent demand, even in non-marginal
urban contexts.

This phenomenon is compounded by a structural oversupply in certain traditional categories,
particularly standard office space and non-specialized residential assets. In these
segments, competition has progressively intensified, compressing rents and eroding
marginal profitability. In the absence of a revision of the asset’s economic function,
purely physical redevelopment interventions tend to produce temporary benefits without
addressing the structural causes of underperformance.

\subsection{Evolution of Demand}

The transformation of real estate demand cannot be reduced to temporary fashions or
short-term cycles. Rather, it is the result of multiple structural dynamics acting
simultaneously and persistently. Ignoring this structural nature leads to superficial
interpretations and redevelopment choices driven by market narratives rather than by
consolidated evidence.

From a demographic perspective, population ageing and the growth of single-person
households are significantly reshaping housing needs. These trends are accompanied by
increased residential mobility, which shortens the average time horizon of housing choices
and strengthens demand for flexible solutions.

At the same time, transformations in the world of work have profoundly altered the
geography of real estate demand. The diffusion of hybrid working arrangements and the
declining centrality of the traditional office have reduced the relevance of certain
historically privileged locations, while increasing the importance of proximity and
accessibility to everyday services.

A further structural element is the progressive decline of ownership-based demand,
particularly among younger population segments. In many urban contexts, property ownership
is losing centrality in favour of use-oriented models that prioritize flexibility, cost
predictability, and service integration. Within this framework, real estate assets are no
longer perceived merely as physical spaces, but as platforms capable of delivering a
coherent bundle of functions and services.

These dynamics do not operate in isolation; rather, they reinforce one another. Their
combined effect renders increasingly inadequate any redevelopment approach that simply
replicates established uses without questioning their actual capacity to intercept present
and future demand.

\subsection{Implications for Value Creation}

The implications of this context for real estate value creation are profound and require a
reconsideration of traditional paradigms. First, the assumption that an asset should retain
a single, stable use over the long term becomes increasingly untenable. The speed at which
demand conditions change makes this approach particularly fragile, especially for medium-
and large-scale assets.

As a result, functional reconfigurability emerges as a critical value attribute. The ability
to adapt an asset to different uses over time, with contained costs and timeframes, becomes
a key factor in assessing its economic resilience. This reconfigurability is not merely a
design issue, but the outcome of strategic choices made during redevelopment, aimed at
avoiding excessively rigid or hyper-specialized solutions.

In parallel, the economic value of real estate assets tends to shift away from pure rental
income towards managerial capability and the quality of the underlying operating model. In
many cases, the difference between an underperforming asset and a valorizable one lies less
in its physical characteristics than in the economic function it performs and the way in
which it is managed.

Within this context, redevelopment cannot be understood as a simple exercise in physical
transformation. Rather, it represents a process of strategic repositioning, in which the
choice of use, the operating model, and the governance structure precede and guide design
decisions. It is on this ground that long-term value creation is ultimately determined.

\section{Assets with Unrealized Potential}

\subsection{Operational Definition}

In the context of real estate redevelopment, the expression ``assets with unrealized
potential'' is often used in a generic or imprecise manner. In this document, it assumes
a specific operational definition: assets that exhibit economic performance below their
theoretical potential, despite not necessarily being degraded, unusable, or located in
marginal contexts. Their underperformance derives primarily from a misalignment between
their current intended use and the demand actually present in the market.

It is important to emphasize that unrealized potential does not coincide with abandonment
or physical deterioration. On the contrary, many assets in good structural and technical
condition are underperforming precisely because they continue to be used according to
economic functions that are no longer capable of intercepting solvent demand. In such
cases, maintenance or refurbishment interventions do not resolve the underlying issue,
which is functional and strategic in nature.

The concept of unrealized potential therefore implies a change in perspective: attention
shifts from the asset as a physical object to the asset as an economic instrument. The
central question is not whether the building is technically recoverable, but whether the
function it performs is still coherent with prevailing market conditions.

\subsection{Recurring Typologies}

Assets with unrealized potential tend to appear in recurring forms, reflecting structural
transformations in real estate demand. A first significant category is represented by
obsolete office buildings, often designed for centralized and rigid work models that have
been partially superseded by the diffusion of hybrid working arrangements and by a decline
in demand for traditional office space. In these cases, the loss of attractiveness depends
less on construction quality than on functional inadequacy.

Another frequent typology concerns former industrial or light-logistics assets, dismissed
or underutilized as a result of delocalization processes or production reorganization.
These assets, while often featuring interesting spatial characteristics, are difficult to
enhance if maintained in their original intended use.

Oversized residential assets also fall among the most common typologies of unrealized
potential. In urban contexts characterized by increasingly smaller households and greater
residential mobility, properties designed for large family units struggle to find an
efficient market positioning, especially when not supported by services or appropriate
management models.

Finally, a significant share of unrealized potential is found in underperforming hotel
assets. Facilities that are no longer able to compete in terms of experience, flexibility,
or service quality tend to progressively lose market share, even in tourist-attractive
areas. In these cases, the issue is not demand per se, but the inability of the traditional
format to respond to evolving customer expectations.

These typologies share a common characteristic: the primary limitation does not lie in the
asset as such, but in the economic function it performs. It is this function, rather than
the physical structure, that determines the asset’s capacity to generate value.

\subsection{Common Valuation Errors}

The presence of unrealized potential is often accompanied by systematic errors in the
valuation phase, which contribute to perpetuating underperformance rather than resolving
it. One of the most frequent errors consists in confusing cadastral value or book value
with the economic value of the asset, thereby neglecting its actual capacity to generate
future cash flows. This confusion leads to an overestimation of asset value and an
underestimation of the need for strategic repositioning.

A second error concerns the underestimation of use-transition costs. Changes in intended
use often entail regulatory adjustments, non-negligible authorization timelines, and
technical-system reconfigurations that significantly affect both costs and implementation
timeframes. Treating these elements as marginal or deferrable results in overly optimistic
valuations that rarely withstand the test of execution.

Finally, a particularly critical error lies in ignoring the management factor. Many
valorization models implicitly assume that asset management is comparable to traditional
leasing, even when the intended use requires high operational intensity and specific
capabilities. In such cases, the failure to consider management as a structural variable
produces significant gaps between expected and actual performance.

Taken together, these errors generate valuations that appear robust but are intrinsically
fragile, with their criticalities tending to emerge only in the later stages of the
project, when the scope for correction has already become limited.

\section{Alternative Uses: A Critical Overview}

The growing diffusion of alternative uses is often presented as an automatic response to
the crisis of traditional real estate typologies. In reality, such solutions do not
constitute a homogeneous set nor a universally valid answer. Rather, they represent a
repertoire of functional models that can generate value only if applied selectively and
coherently with the context, the scale of the asset, and the available managerial
capabilities.

A critical reading of the main alternative uses makes it possible to move beyond
trend-driven narratives and to refocus attention on the structural conditions that
determine their actual sustainability.

\subsection{Advanced Living}

The term ``advanced living'' refers to a set of residential solutions that go beyond the
traditional model of housing as an exclusively individual and purely residential space.
Within these models, the asset assumes the role of an integrated living platform, in which
services, shared spaces, and active management become central elements of the value
proposition.

Student housing represents one of the most established examples of this approach. It is
oriented toward temporary communities and is based on a delicate balance between economic
accessibility, service quality, and proximity to university hubs. Its sustainability
depends less on construction quality than on the ability to intercept a specific demand
and to efficiently manage a high-turnover population.

Microliving follows a similar logic, but targets a broader audience, often composed of
young professionals, mobile workers, or single individuals. In this case, the reduction
of private space is offset by location and the availability of shared services. Here as
well, value does not reside in space per se, but in the ability to offer a flexible housing
solution consistent with highly mobile lifestyles.

Multifamily and Build-to-Rent introduce an additional dimension related to the
professional management of residential assets. These models focus on cash-flow stability
and process standardization, but require adequate scale and a solid management structure.
Their effectiveness depends on the ability to reduce vacancy risk and to maintain a
sustainable balance between provided services and rent levels.

Senior housing, in its ``light'' version, targets a self-sufficient population oriented
toward safe residential environments equipped with services. In this case as well, the
managerial component is decisive: the asset must respond to specific needs that go beyond
mere accommodation, without drifting into complex care-oriented models.

What these advanced living configurations share is a strong dependence on the quality of
the experience offered and on managerial capability. In the absence of these elements,
the risk is to simply replicate a reduced or refurbished residential product without
capturing its true economic potential.

\subsection{Flexible Work}

In the work segment, alternative uses directly reflect transformations in employment and
organizational models. The progressive decline in the centrality of the traditional office
has opened space for more flexible solutions, but also for uses that are more exposed to
operational risk.

Urban coworking represents the most visible form of this transformation. It is a
highly management-intensive model, in which profitability depends on the ability to
attract and retain an active user community. The mere provision of space is insufficient:
value emerges from the combination of services, networking, and dynamic supply management.
As a result, operational risk is high and difficult to sustain in the absence of specific
capabilities.

Neighborhood nearworking constitutes a lighter and more localized articulation of
flexible work. In this case, the value proposition is linked to proximity and the reduction
of commuting, rather than to the creation of complex professional ecosystems. This model
tends to exhibit lower managerial intensity, but also a more limited demand catchment,
which must be carefully verified.

In both configurations, value does not lie in the physical space, but in the ability to
aggregate a real community and to respond to specific work-related needs. Without this
component, the risk is to turn the asset into an underutilized container exposed to rapid
cycles of obsolescence.

\subsection{Evolving Hospitality}

The evolution of hospitality reflects a demand increasingly oriented toward hybrid,
flexible, and de-seasonalized experiences. Emerging models move beyond the traditional
distinction between tourist accommodation and temporary residence, introducing solutions
that integrate work, social interaction, and services.

Hosteling, for example, is oriented toward social interaction and a young, mobile public,
for whom the shared experience is an integral part of perceived value. The sustainability
of this model depends on the ability to maintain a balance between economic accessibility
and offer quality, as well as on careful management of community dynamics.

Hybrid hospitality combines temporary residential functions, workspaces, and services,
responding to the needs of an increasingly fragmented demand. This approach offers greater
flexibility of use, but introduces significant managerial complexity, requiring
multidisciplinary capabilities and clear governance.

Mixed hospitality solutions further extend this logic by integrating multiple functions
within the same asset. Their effectiveness, however, is strictly linked to the ability to
coordinate different components and to avoid inefficient overlaps.

Overall, evolving hospitality offers interesting opportunities, but is characterized by a
high operational-risk profile. Without adequate management, complexity tends to rapidly
erode potential value.

\subsection{Mixed-Use}

Mixed-use is often presented as a universal response to urban complexity and the need for
functional flexibility. In reality, it represents an effective solution only under
well-defined conditions.

For a mixed-use project to generate value, a sufficient critical mass must exist to
support the integrated functions. Moreover, functions must be genuinely synergistic,
capable of reinforcing one another rather than competing for the same resources. Finally,
the complexity introduced by functional integration must be compensated by tangible
added value, in terms of economic resilience or overall attractiveness.

In the absence of these conditions, mixed-use risks becoming a conceptual trap. The
multiplication of functions increases technical and managerial complexity without
producing proportional benefits, exposing the project to delays, inefficiencies, and
management difficulties. In such cases, mixed-use does not solve the asset’s problems, but
amplifies them.

\section{Evaluation Methodology}

The proposed framework is real-options-aware rather than option-pricing-based. 
Its objective is not to estimate the monetary value of embedded options, but to 
preserve option value by controlling irreversibility at the decision stage. 
This is achieved through early exclusion of structurally incoherent uses, 
penalisation of excessive complexity, and prioritisation of reversible and 
time-consistent strategies.

\subsection*{A Decision-Making Framework for Use Selection}

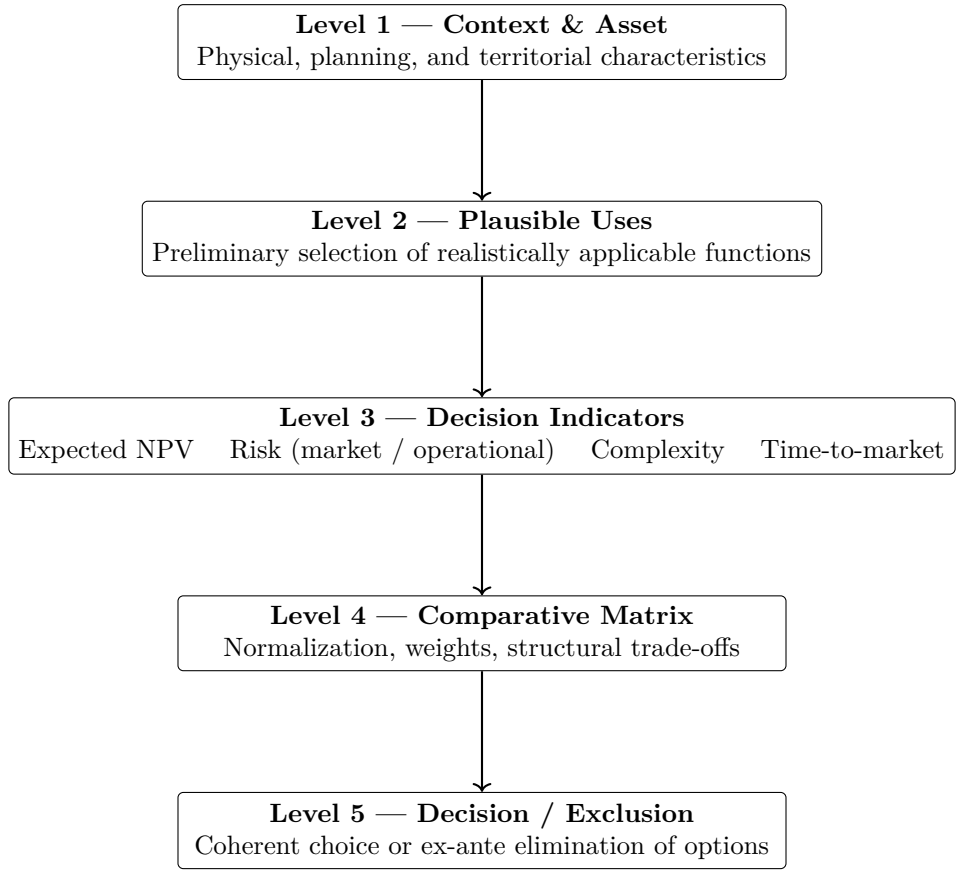
\begin{figure}[h]
	\centering
	\begin{tikzpicture}[
		node distance=1.6cm,
		every node/.style={
			draw,
			rectangle,
			rounded corners=2pt,
			align=center,
			minimum width=8cm,
			minimum height=0.9cm,
			font=\small
		},
		arrow/.style={
			->,
			thick
		}
		]
		
		\node (l1) {%
			\textbf{Level 1 — Context \& Asset}\\
			Physical, planning, and territorial characteristics
		};
		
		\node (l2) [below=of l1] {%
			\textbf{Level 2 — Plausible Uses}\\
			Preliminary selection of realistically applicable functions
		};
		
		\node (l3) [below=of l2] {%
			\textbf{Level 3 — Decision Indicators}\\
			Expected NPV \quad Risk (market / operational) \quad Complexity \quad Time-to-market
		};
		
		\node (l4) [below=of l3] {%
			\textbf{Level 4 — Comparative Matrix}\\
			Normalization, weights, structural trade-offs
		};
		
		\node (l5) [below=of l4] {%
			\textbf{Level 5 — Decision / Exclusion}\\
			Coherent choice or ex-ante elimination of options
		};
		
		\draw[arrow] (l1) -- (l2);
		\draw[arrow] (l2) -- (l3);
		\draw[arrow] (l3) -- (l4);
		\draw[arrow] (l4) -- (l5);
		
	\end{tikzpicture}
	\caption{Logical architecture of the decision-making framework for use selection. The process is structured in sequential levels, with progressive reduction of optionality and control of irreversibility.}
	\label{fig:framework-architecture}
\end{figure}

\subsection{Why ROI Alone Is Not Sufficient}

In real estate redevelopment processes, the use of synthetic indicators such as ROI or
gross yield remains widespread. However, these indicators are conceptually inadequate
when used as \textit{ex-ante} decision-making tools, particularly in the phase of use
selection. Their main limitation lies in their ability to describe a final outcome
without adequately representing the path leading to that outcome.

First, ROI does not explicitly incorporate the time factor. Two projects characterized
by the same percentage return may exhibit profoundly different temporal profiles, with
cash flows anticipated or delayed in ways that produce opposite effects on the present
value of the investment. In the absence of this information, comparisons between
alternatives become misleading.

Second, synthetic return indicators do not distinguish between market risk and
operational risk. A high return may derive either from structurally volatile demand or
from a complex management model highly dependent on execution quality. These are risks of
a different nature, requiring distinct capabilities, tools, and tolerance levels, yet
they are improperly compressed into a single figure.

A further limitation concerns the failure to account for transformation complexity.
Indirect costs, authorization timelines, planning rigidities, and conversion burdens
have a substantial impact on a project’s actual feasibility, but are rarely incorporated
systematically into preliminary yield- or ROI-based assessments. This leads to an
underestimation of required capital and an overestimation of speed to income generation.

Finally, the use of ROI encourages inappropriate comparisons between non-homogeneous
assets. Comparing a coworking space, a Build-to-Rent building, and a student housing
asset on the basis of the same percentage indicator may lead to decisions that are
formally correct but strategically flawed, as they ignore structural differences in
terms of management, risk, and reversibility.

From these considerations follows a clear operational conclusion: ROI may represent a
useful final outcome, but it cannot serve as the tool for selecting intended use in the
initial phase of a redevelopment project.

\subsection{Objective of the Methodology}

The methodology proposed in this paper has a precise and circumscribed objective:
to select the intended use that maximizes expected economic value in a manner compatible
with the risk profile, manageable complexity, and time horizon of the investor or owner.

It does not aim to identify the ``most profitable use in absolute terms,'' nor to apply
formats considered successful in other contexts in a mechanical way. Rather, the
objective is to identify the use that is most coherent with the specific context of the
asset and with the entity responsible for its development and management. From this
perspective, systemic coherence becomes a value criterion at least as important as
expected return.

\subsection{Guiding Principles of the Framework}

The framework is based on a set of operational principles that guide its structure and
define its scope of application. The first principle is comparability: all intended uses
are evaluated on normalized scales in order to enable \textit{ex-ante} comparisons among
alternatives realistically applicable to the same asset.

A second fundamental principle is risk decomposition. Risk is not treated as a unitary
variable, but is explicitly distinguished between market risk and operational risk,
avoiding improper compensation between phenomena of a different nature. This distinction
allows one to understand not only how risky a project is, but also why it is risky.

The third principle concerns the centrality of execution. Managerial capability is not
considered a residual variable or a phase following the project, but a structural factor
that directly affects the economic sustainability of the operation.

The framework also adopts a position of typological neutrality. No intended use is
considered superior by definition: value emerges only from alignment between use,
context, and execution capability.

Finally, the model is conceived as adaptable. Criteria and weightings can be modulated
according to the investor profile, the scale of the operation, and the territorial
context, without compromising overall methodological coherence.

\subsection{Indicators Used}

The methodology is based on six main indicators, organized so as to cover the essential
dimensions of the decision-making process.

Economic value is represented by expected Net Present Value. NPV is not used as a point
estimate, but as a comparative indicator capable of integrating return and time and of
penalizing projects characterized by excessively delayed paybacks. Evaluation is carried
out on a relative scale, using conservative assumptions with respect to rents, occupancy
rates, and operating costs.

Risk is articulated into two distinct components. Market risk measures the volatility and
uncertainty of demand for the intended use considered, taking into account segment
cyclicality, price elasticity, and dependence on exogenous factors. Operational risk, by
contrast, measures the probability that the project will fail to achieve expected results
due to execution issues, even in the presence of demand. It depends on managerial
complexity, the quality of required management, pricing structure, and user churn.

The third dimension concerns complexity and time. Technical complexity includes aspects
related to the physical and regulatory transformation of the asset, such as change of
use, system upgrades, layout reconfiguration, and compliance with specific regulatory
requirements. Managerial complexity, instead, concerns the number and intensity of
processes to be managed, as well as dependence on support systems such as IT, marketing,
and customer care. Alongside these stands time-to-market, which measures the time required
to obtain authorizations, complete the intervention, and reach stable income generation.

\subsection{Weighting of Indicators}

In the standard configuration, weights are assigned so as to reflect a rational,
non-speculative investor profile. Economic value, expressed through expected NPV,
receives the highest weight, while risk and complexity are balanced so as to prevent
theoretically high returns from compensating unsustainable levels of risk or complexity.

This weighting structure is not rigid and can be adapted to different profiles, such as
opportunistic investors, core funds, or non-professional owners. What remains invariant
is the underlying logic, which requires that trade-offs between value, risk, and
execution capability be made explicit.

Weights are decision-profile parameters, not estimated coefficients: they encode explicit
preferences over value, risk, and complexity rather than statistical relationships inferred
from data.

\subsection{Methodological Outputs}

Application of the framework produces a normalized comparative matrix that allows the
ordering of alternative intended uses, a ranking coherent with the adopted criteria, and
the identification of structural trade-offs among the considered options. A central
aspect is the ability to define \textit{ex-ante} exclusion criteria, eliminating appealing
but incoherent alternatives before committing time and capital to more in-depth analyses.

\subsection{Declared Limitations of the Model}

It is important to clarify that the framework does not replace technical due diligence,
planning studies, or an executable business plan. Its role is different and
complementary: to drastically reduce the risk of selecting the wrong intended use in the
initial phase of the redevelopment process, when errors are less costly but more
frequent.

\section{Comparative Decision Matrix}
\subsection*{Construction, Interpretation, and Operational Use}

\subsection{Purpose of the Decision Matrix}

The decision matrix represents the operational core of the framework proposed in this
paper. Its function is to enable the \textit{ex-ante} selection of the most coherent
intended use for a given asset, prior to initiating costly feasibility studies or engaging
in authorization processes that introduce elements of irreversibility.

The matrix is not conceived as a marketing tool, nor as a simplified version of a business
plan. Rather, it acts as a strategic filter, reducing the set of theoretically possible
options to those that are realistically executable, given the characteristics of the
asset, the context, and the decision-maker. Its value lies precisely in its ability to
prevent premature choices that, once crystallized, become difficult to correct.

\begin{center}
	\fbox{%
		\begin{minipage}{0.9\linewidth}
			\small
			\textbf{Methodological clarification.}
			The primary function of the decision matrix is exclusion, not ranking; ranking is secondary and contingent on the adopted decision profile.
	\end{minipage}}
\end{center}

\subsection{Logic of Matrix Construction}

The matrix is designed to compare alternative intended uses applicable to the same asset,
evaluating them through a coherent set of homogeneous and normalized indicators. A central
element of the methodology is the definition of the unit of analysis. The reference unit
is not the asset considered in the abstract, but the combination of asset and intended
use. In other words, what is evaluated is not a functional typology in general, but the
specific impact that a given use would have on a concrete asset.

This approach avoids a frequent error in decision-making processes, namely the abstract
evaluation of intended uses without considering the operational, technical, and
managerial implications they entail when applied to a specific asset.

The structure of the matrix reflects this logic. Each row represents an alternative
intended use, while each column corresponds to a relevant decision indicator. The
indicators employed are those defined in the methodological section and cover the
dimensions of economic value, risk, complexity, and time-to-income. Values are expressed
on a relative scale from one to five, where extreme scores represent the minimum and
maximum relative positions among the considered alternatives.

\subsection{Score Normalization}

Scores assigned within the matrix do not represent absolute values, but relative
positions. This methodological choice is intentional and responds to specific operational
needs. In the preliminary phases of a redevelopment project, available information is
inevitably incomplete and characterized by a high degree of uncertainty. In this context,
the objective is not to estimate final outcomes with precision, but to order options in a
coherent manner.

Normalization allows comparability to be prioritized over point precision. For each
indicator, the alternative with the lowest relative value and the one with the highest
relative value are identified; the remaining options are then positioned coherently along
the scale. This approach enables consistent comparisons, incremental updates of the
matrix as new information emerges, and adaptation of the model to different contexts
without compromising its logical structure.

\subsection{Indicator Aggregation}

To render the matrix an effectively usable tool, elementary indicators are aggregated
into synthetic indices that reflect the main decision dimensions. Economic value is
represented by expected NPV, considered as an autonomous indicator. Overall risk derives
from the integration of market risk and operational risk, while overall complexity
combines technical complexity, managerial complexity, and time-to-income, appropriately
treated as a penalizing factor when excessively long.

Aggregation leads to the definition of an attractiveness index, constructed as a weighted
combination of the three main dimensions. Within this index, economic value contributes
positively, while risk and complexity act as penalizing factors. The adoption of a
subtractive rather than additive structure is deliberate: it makes explicit the trade-offs
among the considered dimensions and prevents high value in one dimension from implicitly
compensating unsustainable levels of risk or complexity.

\subsection{Correct Interpretation of the Matrix}

Interpreting matrix results requires care. A high attractiveness score indicates that the
considered intended use is coherent with the analyzed context, that assumed risks are
proportionate to expected value, and that complexity is manageable for the referenced
decision profile. It does not, however, indicate certainty of success, guaranteed returns,
or absence of risk.

Likewise, a low score does not necessarily signal a calculation error, but rather
highlights a structural misalignment. In such cases, risk or complexity outweigh expected
value, rendering the intended use poorly aligned with prevailing conditions. In many
redevelopment projects, the most useful outcome of the matrix is not the confirmation of
an option, but the timely exclusion of appealing yet inappropriate alternatives.

\subsection{The Matrix as an Exclusion Tool}

One of the most effective uses of the decision matrix lies in its role as an exclusion
tool. The methodology is designed to identify conditions under which an intended use
should be discarded at a preliminary stage. This occurs, for example, when operational
risk is high in the absence of a qualified operator, when technical complexity is
incompatible with available capital, or when time-to-income is inconsistent with the
investor’s time horizon. Even an apparently high NPV may constitute a warning signal if
based on excessively aggressive assumptions.

In such cases, the matrix allows the decision-making process to be halted before
significant resources are committed, thereby reducing the cost of error.

\subsection{Adaptation to Different Decision Profiles}

The matrix is not a rigid instrument. Given the same underlying data, it may produce
different outcomes depending on the adopted weightings. A non-professional owner will
tend to prioritize complexity reduction and strongly penalize operational risk. An
opportunistic investor may assign greater weight to economic value, accepting higher
volatility. An institutional investor, finally, will orient toward stability,
scalability, and standardized management models, penalizing highly operationally
intensive solutions.

This adaptability allows the matrix to remain coherent while the decision-maker profile
changes, making it a flexible yet methodologically robust tool.

\subsection{Limitations of the Matrix}

It is essential to clarify the limits of the decision matrix. It does not replace a
detailed business plan, does not resolve planning issues, nor does it eliminate the
intrinsic risk of a real estate project. Its role is different and complementary: to avoid
the most costly error in real estate redevelopment, namely the initial selection of a
structurally incorrect intended use.

In this sense, the matrix does not promise optimal solutions, but offers a disciplined
method for reducing uncertainty and improving decision quality in the phases where
decisions have the greatest impact.

\section{Use-Selection Strategies}
\subsection*{Aligning Value, Risk, and Execution Capability}

\subsection{General Selection Principle}

The decision matrix allows alternative intended uses to be ranked, but it does not replace
strategic judgment. Final selection cannot be delegated to a simple comparison of scores,
as the optimal intended use does not necessarily coincide with the one characterized by
the highest expected economic value. A sustainable choice emerges only when three
fundamental conditions are simultaneously satisfied: expected value must adequately
compensate the risks undertaken, project complexity must be coherent with available
execution capabilities, and implementation time must be compatible with the investor’s
time horizon.

When even one of these conditions is not met, the project incorporates a structural risk
that cannot be eliminated through diversification nor corrected in later phases. In such
cases, the issue is not one of optimization, but of initial framing.

\subsection{Selection Based on Available Capital}

Available capital represents one of the first objective constraints in selecting an
intended use. With limited resources, the realistic objective is not to maximize potential
value, but to reduce exposure to long timelines, technical complexity, and authorization
uncertainty. In this range, priority lies in preserving liquidity and flexibility, favoring
reversible interventions with short time-to-income. Solutions such as microliving,
nearworking, or neighborhood-scale coworking may be coherent only if supported by a
verified local demand catchment. By contrast, capital-intensive projects with long
gestation periods tend to immobilize resources excessively, even when they promise
theoretically high returns.

With an intermediate level of capital, selection can move toward a balance between value
and stability. In this range, it becomes possible to accept moderate complexity and to
build replicable models, provided that actual managerial capabilities are not overstated.
Mid-scale student housing, multifamily, and light forms of hybrid hospitality fall within
this perimeter, whereas highly operationally intensive solutions require particular
caution. In these cases, the primary risk is not capital scarcity, but the gap between the
complexity of the chosen model and the ability to govern it.

When capital is abundant, selection logic shifts further. The objective becomes the
maximization of risk-adjusted value and the capacity to absorb systemic complexity.
Projects such as articulated mixed-use developments, large-scale student housing, or
structured senior housing become plausible only in the presence of solid governance,
qualified operating partners, and reliable planning frameworks. In the absence of these
conditions, capital scale amplifies risk rather than mitigating it.

\subsection{Selection Based on Available Capabilities}

Alongside capital, capabilities represent a discriminating factor that is often
underestimated. For a non-professional owner oriented toward asset preservation and
characterized by low tolerance for uncertainty, selection should favor simple and
standardizable management models. In such cases, solutions such as multifamily, essential
microliving, or evolved forms of traditional leasing are more coherent, whereas
highly operationally intensive models—such as coworking, hospitality, or complex
mixed-use—tend to expose the owner to uncontrollable risks.

A professional investor or operator, by contrast, possesses greater capacity to structure
processes, access external expertise, and manage complexity at the portfolio level. Within
this profile, uses such as student housing, coworking, or hybrid hospitality can be
selected consciously, provided that the decision matrix is interpreted not at the level
of the single project, but as part of an overall diversification strategy.

In the case of institutional investors or structured club deals, long-term horizons, low
cost of capital, and an interest in stability and scalability enable complex models to be
addressed through a systemic approach. In these contexts, management is internalized or
entrusted to specialized operators, transforming complexity from a constraint into a
strategic lever.

\subsection{Selection Based on Urban Context}

Urban context exerts a decisive influence on the sustainability of an intended use. In
dense and central urban areas, high potential demand and the presence of attractors make
solutions such as microliving, student housing, or urban coworking plausible. However,
these contexts also entail high costs and stringent regulatory constraints, with the
recurring risk of overestimating the market’s capacity to absorb high rents and
underestimating authorization timelines.

Semi-central and belt areas often offer the best compromise between accessibility, costs,
and planning flexibility. In these contexts, solutions such as multifamily, nearworking,
and light hybrid hospitality can intercept real demand while benefiting from greater
functional adaptability. The relationship between demand catchment and complexity tends
to be more favorable than in central areas.

In peripheral or regeneration areas, selection requires even greater caution. Here,
demand is often latent and strongly dependent on public policies or infrastructure
investments. Mixed-use projects, senior housing, or initiatives anchored to essential
services may be viable only in the presence of sufficient critical mass. The most frequent
error in these contexts lies in applying ``successful'' urban formats without a
structural demand capable of sustaining them.

\subsection{Selection Based on Time Horizon}

The investor’s time horizon constitutes an additional selection criterion. With a short
horizon, priority lies in reducing time-to-market and maintaining high reversibility,
favoring light and easily convertible solutions. With a medium-term horizon, attention
shifts toward cash-flow stability and progressive optimization of management, making
models such as student housing, multifamily, or hybrid hospitality plausible. Over long
horizons, finally, resilience, adaptability, and long-term asset value come into play,
justifying more complex choices such as Build-to-Rent, senior housing, or structured
mixed-use.

\subsection{Operational Synthesis}

Selecting an intended use is not an exercise in design creativity, but a problem of
systemic coherence. When capital, capabilities, urban context, and time horizon are
aligned, the decision matrix tends to converge toward a clear solution. When this
alignment is absent, no business plan—however sophisticated—can compensate for the
initial error. It is at this stage that the quality of the entire redevelopment project
is determined.

\section{Structural Risks and Failure Patterns}
\subsection*{Why many redevelopment projects fail despite positive business plans}

\subsection{The Risk That Does Not Appear in the Models}

In real estate redevelopment, a large share of project failures appears to be associated not with computational error, but with deficiencies in initial problem framing. Preliminary business plans are often internally coherent: assumptions are mutually consistent, cash flows appear mathematically plausible, and projected returns fall within conventional ranges. However, such internal consistency may coexist with a deeper contextual misalignment between the project and the environment in which it is expected to operate.

Risk becomes structurally relevant when an intended use is formally viable in financial terms, yet misaligned with the operational and institutional context that conditions its real-world performance. In these situations, risk cannot be attributed to a small number of correctable variables, but instead takes on a systemic character. It becomes difficult to diversify or mitigate \emph{ex post} because it is embedded in the initial specification of use and operating model.

\subsection{Execution Risk}

Among the structural sources of fragility identified in redevelopment projects, execution risk frequently emerges as one of the least adequately represented in early-stage evaluations. Execution risk does not concern architectural feasibility, aggregate demand, or financing structure in isolation, but the capacity to operate the selected model in a stable and coherent manner over time.

Empirically, this risk is most visible in projects where management intensity is implicitly assumed rather than explicitly modelled. Formats such as coworking, student housing, hybrid hospitality, or light senior housing require sustained relational, service, and process management in order to maintain occupancy and revenue stability. Where these managerial functions are weakly specified, initial uptake may still occur, but retention and pricing power tend to deteriorate over time, with negative effects on cash flow and asset value.

Observed failure patterns suggest that this dynamic is frequently linked to the assumption that management can be optimised after the asset has been delivered. When the operating model is not defined \emph{ex ante}, however, the project lacks a stable mechanism through which expected revenues can be realised, regardless of design quality or the formal coherence of financial projections.

\subsection{The Risk of Demand Overestimation}

A second recurring source of misalignment concerns the overestimation of effective demand. In preliminary analyses, potential catchment is often conflated with solvent demand, even though the two differ materially. The presence of a large population or a statistically relevant demographic segment does not, by itself, imply the existence of users willing and able to pay for a specific product in a specific location.

This distinction is particularly relevant in segments such as student housing or young professional accommodation. Aggregate enrolment or population figures provide limited information about willingness to pay, locational preferences, or format acceptance. When demand is not segmented with sufficient granularity, absorption and pricing assumptions tend to become overly optimistic.

Typical indicators of this condition include undifferentiated demand estimates, weak reliance on local transaction benchmarks, rental levels materially above market comparables without a differentiated value proposition, and linear absorption trajectories. When such features are present, sensitivity to pricing and occupancy shocks increases, leading to extended break-even periods and downward pressure on asset value.

\subsection{Regulatory and Planning Risk}

Regulatory and planning constraints constitute a further structural determinant of redevelopment performance. Change of use is often treated as an administrative step, yet it directly influences timelines, spatial configuration, compliance costs, and the range of feasible exit options.

Projects that proceed under optimistic regulatory assumptions or incomplete planning verification tend to accumulate latent rigidities. These constraints typically become visible only after design and capital commitments have been made, at which point adjustment costs are high. Empirical evidence from redevelopment practice indicates that planning delays of twelve to eighteen months can materially affect project NPV and financing requirements, with limited scope for recovery through subsequent optimisation.

\subsection{The Risk of Uncompensated Complexity}

Redevelopment inherently introduces technical, organisational, and regulatory complexity. This complexity becomes a source of fragility when it is not offset by a commensurate increase in expected value. Each additional layer of functional integration, service provision, or governance coordination increases fixed costs and reduces operational flexibility.

Failure patterns frequently emerge in projects that combine multiple uses without sufficient critical mass or without demonstrable synergies. In such cases, narrative or design rationales substitute for demand-based justification, leading to elevated coordination burdens and weaker overall performance.

\subsection{The Risk of Insufficient Local Catchment}

Another structural vulnerability concerns formats that lack a sufficiently deep local user base. Even when aligned with broader market trends, such formats may be unsustainable in contexts where demand is thin, highly seasonal, or dependent on exogenous factors.

This condition is especially prevalent in peripheral or regeneration areas, where metropolitan models are sometimes replicated without an underlying catchment capable of supporting them. In these settings, short-term occupancy may be achieved through price concessions or incentives, but long-term revenue stability is difficult to maintain.

\subsection{The Risk of Design Rigidity}

Design decisions also shape long-term risk exposure. Highly specialised spatial configurations constrain future re-use options and amplify obsolescence risk in environments characterised by shifting demand. When functional reversibility is low, the asset becomes more sensitive to changes in market conditions.

From a strategic perspective, the preservation of at least one viable alternative use increases the option value embedded in redevelopment and reduces exposure to irreversible misalignment.

\subsection{Synthesis of Failure Patterns}

Taken together, these patterns indicate that redevelopment failures are most frequently associated with early-stage misalignment rather than with unanticipated external shocks. Execution risk, demand mis-specification, regulatory rigidity, and uncompensated complexity tend to originate in the initial framing of the project.

Making these mechanisms explicit provides the analytical basis for the structured decision processes developed in the following section, which aim to contain irreversibility and improve the robustness of redevelopment outcomes under uncertainty.

\section{Operational Guidelines for Redevelopment}
\subsection*{From Strategic Decision to Controlled Execution}

\subsection{Redevelopment as a Sequential Process}

Real estate redevelopment can be represented analytically as a staged decision process in which successive phases introduce increasing degrees of irreversibility. Rather than a linear construction workflow, redevelopment is more appropriately understood as a sequence of commitments, each of which constrains the feasible option set available at later stages.

When redevelopment is treated primarily as a physical transformation project, two recurrent sources of risk tend to arise: strategic choices may be fixed prematurely, and capital may be committed to phases that have not yet been validated with respect to demand, regulatory feasibility, or operational coherence. Both mechanisms increase exposure to structural misalignment.

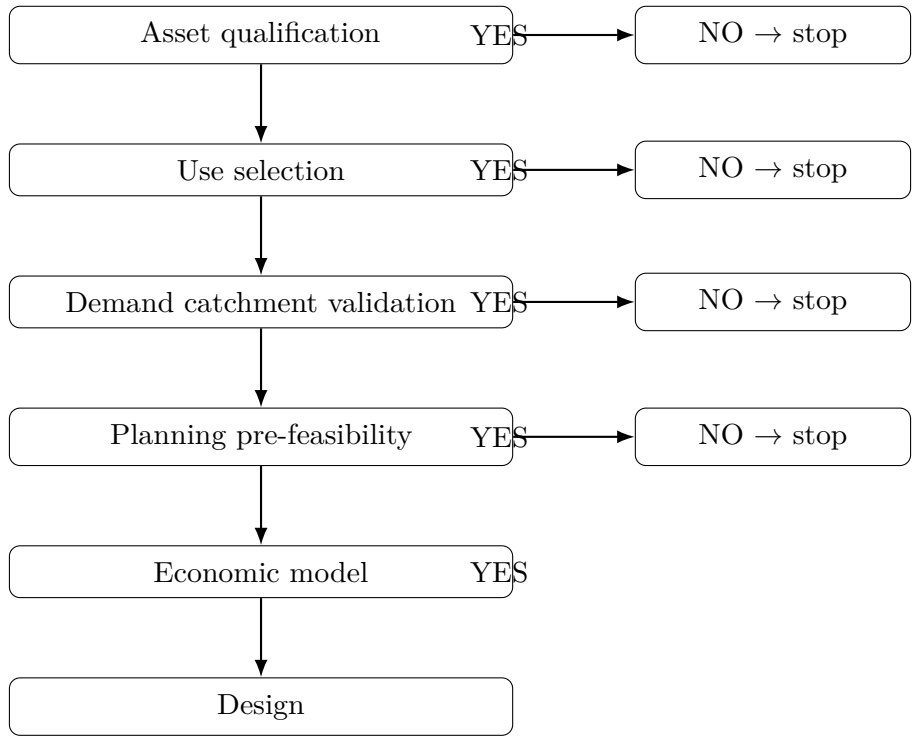
\begin{figure}[h]
	\centering
	\begin{tikzpicture}[
		node distance=1.05cm,
		box/.style={draw, rounded corners, inner sep=6pt, align=center, text width=6.2cm},
		stop/.style={draw, rounded corners, inner sep=6pt, align=center, text width=3.2cm},
		arrow/.style={-Latex, thick}
		]
		
		\node[box] (q)  {Asset qualification};
		\node[box, below=of q] (s) {Use selection};
		\node[box, below=of s] (v) {Demand catchment validation};
		\node[box, below=of v] (p) {Planning pre-feasibility};
		\node[box, below=of p] (m) {Economic model};
		\node[box, below=of m] (d) {Design};
		
		\draw[arrow] (q) -- (s);
		\draw[arrow] (s) -- (v);
		\draw[arrow] (v) -- (p);
		\draw[arrow] (p) -- (m);
		\draw[arrow] (m) -- (d);
		
		\node[stop, right=1.6cm of q] (stopq) {NO $\rightarrow$ stop};
		\node[stop, right=1.6cm of s] (stops) {NO $\rightarrow$ stop};
		\node[stop, right=1.6cm of v] (stopv) {NO $\rightarrow$ stop};
		\node[stop, right=1.6cm of p] (stopp) {NO $\rightarrow$ stop};
		
		\draw[arrow] (q.east) -- (stopq.west);
		\draw[arrow] (s.east) -- (stops.west);
		\draw[arrow] (v.east) -- (stopv.west);
		\draw[arrow] (p.east) -- (stopp.west);
		
		\node[align=left, right=0.2cm of q, xshift=-0.9cm] {YES};
		\node[align=left, right=0.2cm of s, xshift=-0.9cm] {YES};
		\node[align=left, right=0.2cm of v, xshift=-0.9cm] {YES};
		\node[align=left, right=0.2cm of p, xshift=-0.9cm] {YES};
		\node[align=left, right=0.2cm of m, xshift=-0.9cm] {YES};
		
	\end{tikzpicture}
	\caption{Stage-based decision sequence with stop criteria (kill criteria) in preliminary phases}
	\label{fig:decision-sequence}
\end{figure}

The process structure illustrated in Figure~\ref{fig:decision-sequence} is therefore designed to contain early irreversibility, preserve optionality, and provide explicit stopping points before substantial capital and design resources are committed. The objective is not to accelerate redevelopment, but to govern uncertainty in a disciplined manner.

\subsection{The Correct Decision Sequence}

A redevelopment process that seeks to minimise structural risk begins with asset qualification. Before considering how an asset might be redeveloped, it is necessary to determine whether redevelopment is feasible in principle. This assessment concerns legal status, planning constraints, physical scale, accessibility, and the presence of binding technical or regulatory limitations. If these conditions are not satisfied, subsequent analytical effort is unlikely to be economically meaningful.

Once baseline redevelopability has been established, the process enters the strategic use-selection phase. At this stage, the comparative decision matrix serves to reduce the set of theoretically possible uses to a limited number of options that are coherent with the asset and the decision-maker profile. The function of this phase is not confirmation but exclusion: the quality of the outcome depends on the transparency with which implausible alternatives are eliminated.

The next phase concerns validation of the demand catchment. Here the objective is not to estimate abstract potential demand, but to assess the existence of solvent demand under realistic pricing and locational conditions. This typically requires reference to local transaction benchmarks, observed absorption patterns, and consultation with active operators. Reliance on macro-level indicators in place of local evidence is a common source of error at this stage.

Only after demand coherence has been established does it become meaningful to undertake technical and planning pre-feasibility analysis. Change-of-use constraints, compliance requirements, system capacity, and regulatory adaptation costs must be evaluated before financial optimisation, since planning feasibility conditions the entire economic structure of the project.

Preliminary economic modelling then serves to test sustainability, bankability, and consistency with the investor’s time horizon. Models at this stage are necessarily conservative and sensitivity-oriented, with emphasis placed on cash-flow dynamics and Net Present Value rather than on single-point return metrics. Depending on results, the project may proceed, be resized, or be discontinued.

Design and execution represent the final phase, in which capital is committed irreversibly and spatial configurations become fixed. Advancing to this stage without prior validation implies an acceptance of avoidable strategic risk.

\subsection{The Role of Preliminary Checks}

Progression beyond the early phases of redevelopment is conditional on a number of minimum requirements. The intended use must be compatible with asset scale and context; the existence of solvent demand must be supported by local evidence; operational requirements must be consistent with available capabilities; and time-to-income must fall within the decision-maker’s horizon. In addition, at least a preliminary operating model must be specified.

Failure to satisfy any of these conditions indicates that the project has not yet reached a sufficient level of maturity to justify irreversible commitments.

\subsection{Stop Criteria as a Capital-Protection Mechanism}

Redevelopment processes frequently display path dependence: once time and resources have been invested, there is a tendency to continue even when project fundamentals deteriorate. The absence of explicit stopping criteria amplifies this effect.

Analytically, a project may be considered unsuitable for continuation when planning feasibility introduces binding constraints, when demand validation fails, when time-to-income exceeds the investor’s tolerance, or when added complexity is not compensated by higher expected value. Early termination under such conditions represents a rational capital-allocation decision rather than a project failure.

\subsection{Reversibility and Optionality}

Under structural uncertainty, reversibility constitutes an economically relevant attribute. Designs and configurations that preserve alternative use options increase the option value embedded in an asset and reduce exposure to accelerated obsolescence. Excessively specialised layouts, by contrast, convert uncertainty into irreversibility and magnify downside risk.

From a decision-theoretic perspective, the preservation of at least one viable exit or reconfiguration path functions as a form of strategic insurance.

\subsection{Project Governance and Decision Quality}

Redevelopment outcomes are also shaped by governance structures. Clear differentiation between ownership, development, management, and oversight functions supports accountability and reduces coordination failure, particularly in projects characterised by operational complexity.

Documenting assumptions, risk assessments, and phase-gate decisions at each stage enhances traceability and facilitates learning, both within a given project and across successive redevelopment initiatives.

\subsection{Operational Synthesis}

When redevelopment is treated as a staged decision process rather than as a purely technical exercise, uncertainty can be managed through controlled commitment rather than deferred correction. The sequencing logic presented here aims to minimise the number of irreversible decisions taken before sufficient information about demand, feasibility, and operational viability has been obtained.

\section{Illustrative Applications}
\label{sec:illustrative-applications}

This section presents stylised illustrative cases designed to demonstrate how the proposed
framework supports \textit{ex ante} use selection through comparative screening and early
exclusion of structurally incoherent alternatives. The cases are constructed as composites
reflecting recurring professional patterns; they are not intended as empirical validation
nor as complete business plans, but as analytical devices for examining decision logic
under incomplete information.

\subsection{Case 1 --- Semi-central office conversion (large urban market)}
\label{subsec:case1-office}

\textbf{Asset and context.}
The reference asset is a former 1980s office building of approximately
2{,}800~m\textsuperscript{2} GLA, located in a semi-central urban area with good accessibility
but without primary metropolitan hubs. The building is technically serviceable and physically
convertible, yet underperforms as traditional office space in a context characterised by
oversupply and changing work patterns.

The decision-maker profile corresponds to a non-institutional investor with limited tolerance
for prolonged vacancy, moderate capital availability, and no in-house operating platform.

\medskip

\textbf{Plausible intended uses (preliminary set).}
Based on baseline technical and planning plausibility, four alternatives are considered:
urban coworking; light mixed-use; multifamily/build-to-rent; and microliving.

\medskip

\textbf{Application of the decision matrix (summary).}
Relative scores (1--5) are assigned for expected NPV, market risk, operational risk, technical
complexity, managerial complexity, and time-to-income. Two options are characterised by
structural misalignment:

\begin{itemize}
	\item \textbf{Urban coworking} exhibits elevated operational risk and managerial complexity
	that are not offset by asset scale or by the depth of the local demand catchment. Risk is
	primarily execution-related rather than construction-related.
	\item \textbf{Light mixed-use} introduces technical and governance complexity that is
	disproportionate to the scale of intervention, without a corresponding increase in
	expected value.
\end{itemize}

These options are therefore excluded \textit{ex ante}.

\medskip

\textbf{Resulting shortlist and next step.}
The remaining options, \textbf{multifamily/build-to-rent} and \textbf{microliving}, display
more balanced profiles in terms of expected value, complexity, and time-to-income. The
framework therefore directs subsequent analysis toward local demand validation for these
alternatives only.

\medskip

\textbf{Worked example (auditable illustration).}
Table~\ref{tab:case1-worked-matrix} reports one numerical instantiation of the matrix for this
case. Values are expressed on a relative 1--5 scale and serve to support comparative screening
rather than return forecasting.

\begin{table}[h]
	\centering
	\small
	\setlength{\tabcolsep}{3.5pt}
	\renewcommand{\arraystretch}{1.15}
	\caption{Worked comparative matrix for Case~1 (illustrative, auditable example).}
	\label{tab:case1-worked-matrix}
	\begin{tabular}{lccccccccc}
		\toprule
		\textbf{Use} &
		\textbf{NPV} &
		\textbf{$R_m$} &
		\textbf{$R_o$} &
		\textbf{$R$} &
		\textbf{$C_t$} &
		\textbf{$C_g$} &
		\textbf{$T$} &
		\textbf{$C$} &
		\textbf{$A$} \\
		\midrule
		Urban coworking    & 4 & 3 & 5 & 4.0 & 2 & 5 & 3 & 3.4 & -0.62 \\
		Light mixed-use    & 4 & 3 & 4 & 3.5 & 5 & 4 & 1 & 3.6 & -0.73 \\
		Multifamily / BTR  & 3 & 2 & 2 & 2.0 & 3 & 2 & 4 & 2.7 & \phantom{-}0.21 \\
		Microliving        & 3 & 3 & 2 & 2.5 & 2 & 2 & 4 & 2.4 & \phantom{-}0.33 \\
		\bottomrule
	\end{tabular}
	
	\medskip
	\footnotesize
	\justifying
	\textbf{Notes.}
	Overall risk $R = 0.5R_m + 0.5R_o$.
	Overall complexity $C = 0.3C_t + 0.4C_g + 0.3T$.
	Attractiveness $A = 0.40\cdot \text{NPV} - 0.30\cdot R - 0.30\cdot C$.
	Negative $A$ indicates \textit{ex ante} exclusion under the adopted decision profile.
\end{table}

\medskip

The worked example reproduces the qualitative screening outcome: coworking is primarily
penalised by operational and managerial exposure, while mixed-use is penalised by
uncompensated technical and governance complexity. Multifamily and microliving exhibit
comparatively coherent profiles and therefore proceed to the next validation stage.

\medskip

\textbf{Analytical implication.}
In large urban markets, the dominant source of fragility is not the absence of demand, but
the adoption of formats whose operational exposure and complexity reduce option value at
the scale considered. The matrix operates as a structural filter, preventing premature
over-complexification.

\subsection{Case 2 --- Underperforming hotel repositioning (medium-sized city)}
\label{subsec:case2-hotel}

\textbf{Asset and context.}
The asset is a former hotel of approximately 2{,}500~m\textsuperscript{2} GLA located near the
historic centre of a medium-sized city (50{,}000--150{,}000 inhabitants), characterised by
moderate and seasonal tourism and the presence of a small-to-medium university. The building
is physically adequate but economically fragile in its current use.

The owner profile corresponds to a private decision-maker with limited managerial bandwidth
and a preference for income stability.

\medskip

\textbf{Plausible intended uses (preliminary set).}
The alternatives considered at the \textit{ex ante} stage are: hybrid hospitality; student
housing; light senior housing; coworking plus events; and traditional residential use.

\medskip

\textbf{Application of the decision matrix (summary).}
The comparative screening highlights two sources of misalignment:

\begin{itemize}
	\item \textbf{Coworking plus events} is characterised by high operational volatility and
	dependence on continuous activation of demand, resulting in elevated execution risk.
	\item \textbf{Hybrid hospitality} displays elevated complexity and exposure to seasonal
	demand, with value strongly contingent on managerial performance and specialised layouts.
\end{itemize}

These options are therefore excluded \textit{ex ante}.

\medskip

\textbf{Resulting shortlist and next step.}
The matrix identifies \textbf{light senior housing} as the most coherent option, with
\textbf{small-scale student housing} as a secondary candidate. Both are associated with more
stable demand and longer average stays, motivating further validation of pricing, regulatory
conditions, and operating models.

\medskip

\textbf{Analytical implication.}
In medium-sized cities, redevelopment risk is strongly influenced by the depth and stability
of the local catchment. Screening therefore tends to favour uses anchored to structural needs
over formats whose performance depends on volatile utilisation.

\subsection{Case 3 --- Redevelopment in inner areas (small municipality)}
\label{subsec:case3-inner}

\textbf{Asset and context.}
The reference asset is a former public or semi-public building of approximately
1{,}600--2{,}000~m\textsuperscript{2} GLA located in a small municipality or inner area
characterised by ageing demographics, limited accessibility, and weak private real estate
dynamics.

The decision-maker is typically a local public body, a small private owner, or a mixed
public--private vehicle operating under capital constraints and long-term service
objectives.

\medskip

\textbf{Plausible intended uses (preliminary set).}
The alternatives considered are: light senior housing; assisted or semi-assisted housing;
community-oriented mixed-use; free-market residential; and rural coworking.

\medskip

\textbf{Application of the decision matrix (summary).}
Three options exhibit structural misalignment:

\begin{itemize}
	\item \textbf{Rural coworking} is constrained by insufficient effective demand and high
	operational fragility.
	\item \textbf{Free-market residential} is characterised by weak expected NPV and extended
	absorption periods.
	\item \textbf{Community mixed-use} introduces coordination and governance complexity that
	exceeds local execution capacity.
\end{itemize}

These options are therefore excluded \textit{ex ante}.

\medskip

\textbf{Resulting shortlist and next step.}
The matrix identifies \textbf{light senior housing} (or a closely related assisted-living
configuration) as the only option consistent with local demand structure and execution
capability. Subsequent analysis focuses on service model design, regulatory compatibility,
and institutional partnerships rather than on market-driven pricing optimisation.

\medskip

\textbf{Analytical implication.}
In inner areas, redevelopment value is primarily associated with functional anchoring to
essential services rather than with market-based demand expansion. The decision matrix
functions as a reality filter, excluding formats whose apparent flexibility masks structural
demand insufficiency.

\medskip

Numerical instantiations analogous to Table~\ref{tab:case1-worked-matrix} can be produced for
Cases~2 and~3 using the same definitions (Appendix~D); they are omitted here to preserve
analytical focus on decision logic rather than on numerical detail.

\section{Conclusions}
\subsection*{Redevelopment Does Not Mean Transformation: It Means Making the Right Decisions}

\subsection{Summary of Findings}

This paper has framed real estate redevelopment not primarily as an architectural or financial optimisation problem, but as a decision-making problem under conditions of irreversibility and structural uncertainty. The analysis suggests that a substantial share of redevelopment risk is introduced in early project phases, when the intended use is selected, rather than during design or execution.

The underperformance observed in a significant portion of the existing real estate stock appears, in many cases, to be structural rather than contingent. Assets frequently underperform not because of physical inadequacy or an absolute lack of demand, but because of misalignment between expected economic value, risk exposure, and the execution capabilities of those who govern them. In this context, reliance on synthetic indicators such as ROI or yield provides an incomplete basis for initial decision-making, as these measures do not adequately capture operational complexity or the temporal structure of cash flows.

From this follows a central implication: the initial choice of intended use exerts a disproportionate influence on overall project risk and long-term performance. When this choice is poorly aligned with context and capabilities, the project incorporates structural risks that are difficult to correct \emph{ex post}.

\subsection{The Required Paradigm Shift}

Effective real estate redevelopment requires a shift in perspective from projects to decisions. Design should be understood not as the starting point of the process, but as the outcome of a prior strategic selection. Sound redevelopment projects originate from choices that are made before irreversible commitments are crystallised.

A second shift concerns the move from maximising theoretical returns to ensuring systemic coherence. Value is not necessarily maximised by pursuing the use with the highest potential upside, but by selecting the option most consistent with asset scale, territorial context, available capabilities, and the investor’s time horizon. Where this alignment is weak, even apparently attractive returns tend to prove fragile.

A further shift involves the transition from typologies to functions. Traditional categories such as offices, residential, and hospitality are increasingly inadequate descriptors in environments characterised by hybrid models and management-intensive operations. What matters is less the nominal category of an asset than the concrete economic and organisational function it performs.

Finally, a logic of early irreversibility must give way to one of optionality. Preserving the ability to adapt or reconfigure uses over time should be regarded not as indecision, but as a strategic response to structural uncertainty.

\subsection{The Role of the Decision Matrix}

Within this paradigm shift, the proposed decision matrix plays a specific and deliberately circumscribed role. It is not intended to provide definitive answers or to replace strategic judgment, but to make explicit the trade-offs that are often implicit in financial models. By enabling \emph{ex ante} comparison among realistically applicable alternatives, it helps reduce the likelihood of committing resources to structurally incoherent projects.

Used appropriately, the matrix is less a tool for identifying an objectively ``best'' project than for excluding those that are poorly aligned with context, capabilities, and risk tolerance. In a sector where many failures originate in early framing errors, this exclusion function represents a substantive contribution.

\subsection{Implications for Practitioners}

The framework has implications for multiple categories of practitioners. For asset owners, it highlights that redevelopment is not always the optimal response to underperformance; in some cases, limiting intervention or moderating ambition may better preserve value.

For investors, competitive advantage is less likely to derive from access to fashionable or innovative formats than from the ability to select them appropriately given context and risk profile. Innovation resides primarily in the decision process rather than in the format itself.

For advisors and developers, professional value increasingly shifts from design towards the governance of early-stage decision complexity. Distinctive competence lies in structuring \emph{ex ante} choices rather than optimising outcomes \emph{ex post}.

Important implications also arise for public administrations. The quality of urban transformation depends not only on regulatory control after the fact, but also on the clarity and robustness of the decision processes that precede design. Improving these processes can contribute to reducing conflict, delay, and systemic inefficiency.

\subsection{Final Consideration}

Real estate redevelopment does not fail because it is complex, but because complexity is
often addressed too late, after fundamental choices have already been fixed and
opportunities for correction have narrowed. In this sense, improving the quality of early
decisions constitutes a central mechanism for risk mitigation.

This paper does not propose a catalogue of formats to be replicated. The contribution is
therefore methodological rather than typological: it concerns how redevelopment choices
are made, not which formats are preferred. Its purpose is to provide a method for assessing
when and which solutions are coherent, situating redevelopment decisions within a
rational framework consistent with the structural complexity of contemporary real estate
markets.

\section*{Limitations and scope}

This work is intentionally positioned \emph{upstream} of feasibility studies and
execution-level business planning. The comparative scores used in the matrix are
\emph{relative} and context-dependent: they support ordering and exclusion of alternatives
under uncertainty, but they do not provide point forecasts of returns.

The framework also relies on disciplined judgment in scoring market risk, operational
risk, and managerial complexity. While the quantitative appendix provides a replicable
normalization and aggregation logic, high-quality application still requires local
benchmarks, operator interviews, and conservative assumptions.

Finally, the approach is designed for \emph{use selection} (and early exclusion) and does
not substitute for legal due diligence, planning verification, technical design,
construction risk management, or financing structuring.

\section*{Reproducibility}

To support replication and auditability in professional settings (e.g., investment
committee review), the methodology is designed to be implemented through a standard
evaluation worksheet capturing:
\begin{itemize}
  \item the set of plausible intended uses considered for a given asset;
  \item raw indicator inputs (NPV components, risk proxies, complexity proxies, timelines);
  \item normalization ranges and resulting 1--5 scores;
  \item weight vectors reflecting the decision-maker profile;
  \item the resulting attractiveness ranking and the explicit \emph{ex ante} exclusion rules.
\end{itemize}

\cleardoublepage
\phantomsection
\addcontentsline{toc}{part}{Appendices}
\appendix

\section{Format Summary Table}
\label{app:formats}

This appendix provides a compressed, decision-oriented summary of the main formats used in
the comparative matrix. The purpose is not descriptive completeness, but comparability:
each row is designed to support \emph{ex ante} screening under uncertainty.

\begin{table}[h]
	\centering
	\scriptsize
	\setlength{\tabcolsep}{3pt}
	\renewcommand{\arraystretch}{1.05}
	\caption{Decision-oriented summary of formats (compressed, signal-based).}
	\label{tab:format-summary}
	\begin{tabularx}{\textwidth}{
			>{\RaggedRight\arraybackslash}p{2.3cm}
			>{\RaggedRight\arraybackslash}p{1.9cm}
			>{\RaggedRight\arraybackslash}p{2.7cm}
			>{\RaggedRight\arraybackslash}p{2.4cm}
			>{\RaggedRight\arraybackslash}p{2.1cm}
			>{\RaggedRight\arraybackslash}p{2.6cm}
		}
		\toprule
		\textbf{Format} &
		\textbf{Scale} &
		\textbf{Context} &
		\textbf{Demand} &
		\textbf{Risk} &
		\textbf{Decision signal} \\
		\midrule
		
		Microliving &
		1{,}000--2{,}000 m\textsuperscript{2} &
		Dense / semi-central &
		Singles, young workers &
		Moderate market &
		\textbf{Product coherence}\\
		{\footnotesize Location and unit design matter; avoid densification-only logic.} \\
		\addlinespace
		
		Student housing &
		3{,}000--5{,}000 m\textsuperscript{2} &
		University hubs, transit &
		Out-of-town students &
		Catchment risk &
		\textbf{Verified demand}\\
		{\footnotesize Scalable only if demand is measured, not assumed.} \\
		\addlinespace
		
		Multifamily / BTR &
		2{,}500--4{,}000 m\textsuperscript{2} &
		Consolidated areas &
		Medium--long stays &
		Low--medium &
		\textbf{Cash-flow stability}\\
		{\footnotesize Patrimonial profile; requires standardised operations.} \\
		\addlinespace
		
		Coworking &
		1{,}200--2{,}500 m\textsuperscript{2} &
		Dense urban &
		Freelancers, SMEs &
		High operational &
		\textbf{Execution-intensive}\\
		{\footnotesize Value driven by services and community, not space.} \\
		\addlinespace
		
		Hybrid hospitality &
		$\sim$3{,}000 m\textsuperscript{2} &
		Mixed-demand nodes &
		Temporary users &
		Complexity risk &
		\textbf{Partner-dependent}\\
		{\footnotesize Optionality high; operator choice is decisive.} \\
		\addlinespace
		
		Senior housing (light) &
		3{,}000--6{,}000 m\textsuperscript{2} &
		Accessible, quiet &
		Self-sufficient 65+ &
		Regulatory risk &
		\textbf{Service continuity}\\
		{\footnotesize Requires organisational and relational oversight.} \\
		\addlinespace
		
		Mixed-use &
		6{,}000--8{,}000 m\textsuperscript{2} &
		Large regeneration &
		Multi-segment &
		Systemic risk &
		\textbf{Critical mass}\\
		{\footnotesize Viable only with real synergies.} \\
		
		\bottomrule
	\end{tabularx}
\end{table}

\section{Quantitative Methodology}
\label{app:quant}

This appendix makes explicit and replicable the quantitative criteria adopted in the
decision-support framework presented in the main body (see also the worked example in
Section~\ref{subsec:case1-office}, Table~\ref{tab:case1-worked-matrix}). Its scope is
deliberately limited: it does not replace detailed business plans, technical due diligence,
or planning due diligence, nor does it provide point estimates of value. Its purpose is to:
\begin{itemize}
	\item reduce arbitrariness in preliminary assessments;
	\item enable coherent comparisons among alternative intended uses;
	\item improve transparency and auditability of the \textit{ex ante} selection process.
\end{itemize}
The guiding principle is \emph{decision robustness}, not financial precision.

\subsection{Unit of analysis and general setup}

The unit of analysis is not the property in absolute terms, but the decision pair:
\[
U_i = (A, D_i),
\]
where:
\begin{itemize}
	\item $A$ denotes the real estate asset under consideration;
	\item $D_i$ denotes a specific plausible intended use.
\end{itemize}

All scores and indices are \emph{comparative} and valid only relative to the set of
alternatives considered for the same asset. They should not be transferred to different
contexts without re-normalization.

\subsection{Score normalization}

\subsubsection*{Rationale}

In preliminary phases, data are incomplete, assumptions are necessarily conservative, and
absolute precision is not a realistic objective. For this reason, the framework adopts a
\emph{relative} normalization approach aimed at ordering alternatives without assigning
absolute meaning to estimates.

\subsubsection*{Normalization method (benefit vs.\ penalty indicators)}

Let $X_i$ denote a raw indicator value for alternative $i$ (e.g., NPV or a time measure),
and let $X_{\min}, X_{\max}$ denote the minimum and maximum values observed among the
considered alternatives.

\paragraph{Benefit indicators (higher is better).}
For indicators where larger values are preferable (e.g., expected NPV), the normalized score
on a 1--5 scale is:
\[
S_i(X) = 1 + 4 \cdot \frac{X_i - X_{\min}}{X_{\max} - X_{\min}} .
\]

\paragraph{Penalty indicators (higher is worse).}
For indicators where larger values are worse (e.g., time-to-income, risk proxies, complexity
proxies), the normalized score is inverted:
\[
S_i(X) = 1 + 4 \cdot \frac{X_{\max} - X_i}{X_{\max} - X_{\min}} .
\]
In all cases, higher scores denote a \emph{more favorable} position for the decision-maker
within the considered set.\footnote{If raw data are already elicited directly on a 1--5
	scale by expert judgment, the normalization step is skipped and the elicited score is used.}

\subsection{Economic value: expected NPV}

\subsubsection*{Operational definition}

Expected NPV is defined as the present value of estimated operating cash flows for a given
intended use, net of initial investment:
\[
NPV_i = \sum_{t=1}^{T} \frac{CF_{i,t}}{(1+r)^t} - I_0,
\]
where $CF_{i,t}$ are net operating cash flows, $r$ is a discount rate consistent with the
risk profile, and $I_0$ is the initial investment.

\subsubsection*{Use within the framework}

Within this framework:
\begin{itemize}
	\item NPV is not treated as a point estimate of value;
	\item only the \emph{relative position} among alternatives is used for screening;
	\item conservative assumptions are adopted for rents, occupancy, operating costs, and
	stabilization timing.
\end{itemize}

The normalized NPV score is denoted $\text{NPV}_i$ in
Table~\ref{tab:case1-worked-matrix}.

\subsubsection*{Minimum stress testing}

Each preliminary NPV should be tested against at least:
\begin{itemize}
	\item a 10\% reduction in rents;
	\item a 10\% reduction in occupancy;
	\item a 12-month delay in revenue generation.
\end{itemize}
An NPV that becomes negative under all scenarios indicates structural fragility of the
intended use under the adopted assumptions and should trigger \textit{ex ante} exclusion.

\subsection{Risk: decomposition and measurement}

\subsubsection*{Market risk}

Market risk captures the stability and depth of demand for a given intended use.
Operational proxies may include:
\begin{itemize}
	\item historical volatility of occupancy or absorption;
	\item price elasticity of demand;
	\item segment cyclicality;
	\item revenue seasonality.
\end{itemize}
The resulting normalized score is denoted $R_{m,i}$ (1--5).

\subsubsection*{Operational risk}

Operational risk captures the probability of performance shortfall linked to execution and
management, regardless of the existence of demand. Proxies may include:
\begin{itemize}
	\item share of revenues dependent on active services (vs.\ passive rent);
	\item user churn or contract turnover intensity;
	\item intensity of managerial oversight required;
	\item degree of operating-model standardization (inverse proxy).
\end{itemize}
The resulting normalized score is denoted $R_{o,i}$ (1--5).

\subsubsection*{Overall risk}

Overall risk aggregates market and operational risk as:
\[
R_i = \alpha R_{m,i} + (1-\alpha) R_{o,i},
\]
where $\alpha \in [0,1]$ reflects the decision-maker’s profile (baseline: $\alpha=0.5$).
In Table~\ref{tab:case1-worked-matrix}, the derived value is reported as $R$.

\subsection{Complexity: quantitative articulation}

\subsubsection*{Technical complexity}

Technical complexity concerns physical and regulatory transformation burdens. Proxies may
include:
\begin{itemize}
	\item need for change of use and planning conformity constraints;
	\item number and specificity of regulatory standards to be met;
	\item CAPEX per square meter and MEP/system reconfiguration intensity;
	\item expected duration and uncertainty of permitting processes.
\end{itemize}
The resulting normalized score is denoted $C_{t,i}$ (1--5).

\subsubsection*{Managerial complexity}

Managerial complexity concerns ongoing operation of the intended use. Proxies may include:
\begin{itemize}
	\item FTE required per 1{,}000 m\textsuperscript{2};
	\item number of core processes to manage (operations, marketing, customer care, IT);
	\item degree of service intensity and frequency of user interaction;
	\item dependence on non-automatable activities.
\end{itemize}
The resulting normalized score is denoted $C_{g,i}$ (1--5).

\subsubsection*{Time-to-income (time-to-market component)}

Time-to-income is defined as:
\[
T_i = T_{\mathrm{permits}} + T_{\mathrm{works}} + T_{\mathrm{stabilization}}.
\]
Within the framework, longer durations are penalized. The normalized score is denoted $T_i$
(1--5) and corresponds to the \emph{time-to-income} concept used in the main text.

\subsubsection*{Overall complexity}

Overall complexity aggregates technical complexity, managerial complexity, and time-to-income:
\[
C_i = \beta C_{t,i} + \gamma C_{g,i} + \delta T_i,
\qquad
\beta + \gamma + \delta = 1.
\]
Baseline values are $\beta/\gamma/\delta = 0.3/0.4/0.3$.
In Table~\ref{tab:case1-worked-matrix}, the derived value is reported as $C$.

\subsection{Attractiveness index}

The attractiveness index synthesizes the three main decision dimensions:
\[
A_i = w_V \cdot \text{NPV}_i - w_R \cdot R_i - w_C \cdot C_i,
\]
with baseline weights:
\begin{itemize}
	\item value: $w_V = 0.40$;
	\item risk: $w_R = 0.30$;
	\item complexity: $w_C = 0.30$.
\end{itemize}
The subtractive structure is deliberate: it makes trade-offs explicit and prevents high
value from implicitly compensating unsustainable risk or uncompensated complexity.

\subsubsection*{Operational interpretation}

\begin{itemize}
	\item $A_i > 0$: coherent intended use (screening pass);
	\item $A_i \approx 0$: borderline option (requires stronger local evidence);
	\item $A_i < 0$: \textit{ex ante} exclusion under the adopted decision profile.
\end{itemize}
The index does not predict success; it reduces the probability of structural selection
error and supports disciplined early-stage filtering.

\subsection{Declared limitations of the model}

The quantitative model:
\begin{itemize}
	\item does not eliminate risk;
	\item does not replace execution-level analyses (due diligence, planning feasibility,
	full operating-model definition);
	\item does not produce definitive value estimates.
\end{itemize}
Its usefulness lies in improving the quality of early-stage decisions, when the cost of
error is lowest and optionality is highest.

\end{document}